\documentclass{article}
\usepackage{spconf,amsmath,graphicx}
\usepackage{xcolor}
\usepackage{url}
\usepackage{subfigure}
\usepackage{cite}

\usepackage{array}
\usepackage{mdwmath}
\usepackage{mdwtab}
\usepackage{eqparbox}
\usepackage{multirow}
\usepackage{stmaryrd}
\usepackage{paralist}
\usepackage{dsfont}
\usepackage{verbatim}
\usepackage{calc}
\usepackage{amssymb}
\usepackage{amstext}
\usepackage{amsthm}
\usepackage{multicol}
\usepackage{xcolor}
\usepackage{rotating}

\usepackage{algorithm}
\usepackage{algorithmic}

\usepackage{hyperref}

\usepackage{xcolor}
\usepackage{pgfplots,pgfplotstable}
\usepackage{tikz}
\usepackage{lipsum,adjustbox}
\usetikzlibrary{calc,positioning}
\pgfplotsset{compat=newest}

\definecolor{light-gray}{gray}{0.75}

\newcommand{\smalllskip}[0]{\vspace{1pt}}

\newcommand{\ie}[0]{\textit{i.e.},~}
\newcommand{\eg}[0]{\textit{e.g.},~}

\title{ColorNet - Estimating Colorfulness in Natural Images}
\name{Emin Zerman$^*$, Aakanksha Rana$^*$\thanks{*These authors contributed equally to this work.}, Aljosa Smolic\thanks{This publication has emanated from research conducted with the financial support of Science Foundation Ireland (SFI) under the Grant Number 15/RP/2776. We also gratefully acknowledge the support of NVIDIA Corporation with the donated GPU used for this research.}}
\address{V-SENSE, School of Computer Science, Trinity College Dublin, Dublin, Ireland}
\begin{document}
\ninept
\maketitle
\begin{abstract}

Measuring the colorfulness of a natural or virtual scene is critical for many applications in image processing field ranging from capturing to display. In this paper, we propose the first deep learning-based colorfulness estimation metric. For this purpose, we develop a color rating model which simultaneously learns to extracts the pertinent characteristic color features and the mapping from feature space to the ideal colorfulness scores for a variety of natural colored images. Additionally, we propose to overcome the lack of adequate annotated dataset problem by combining/aligning two publicly available colorfulness databases using the results of a new subjective test which employs a common subset of both databases. Using the obtained subjectively annotated dataset with 180 colored images, we finally demonstrate the efficacy of our proposed model over the traditional methods, both quantitatively and qualitatively.
\end{abstract}
%

\section{Introduction}
\label{sec:intro}

Color is a crucial factor in human visual perception. It affects human behavior and decision processes both in nature and in society, as color conveys pivotal information about the surroundings. Thus, its accurate acquisition and display are necessary for multimedia, entertainment, and image processing systems. 

Within the imaging pipeline, the color or brightness information of the real (or virtual) scene needs to be processed for various reasons such as color grading~\cite{pitie2007automated}, tone-mapping~\cite{mantiuk2008display} for high dynamic range (HDR) scenes, gamut mapping~\cite{sikudova2016gamut}, color correction~\cite{mantiuk2009color}. During the acquisition or the processing, color of the scene can be affected by color cast or colorfulness changes~\cite{hasler2003measuring}. Within the scope of this study, only the colorfulness aspect of the natural images is considered.

Colorfulness is generally defined as the amount, intensity, and saturation of colors in an image~\cite{fairchild2013color}. Understanding and estimating colorfulness is quintessential for a variety of applications, \eg HDR tone-mapping~\cite{mantiuk2009color, reinhard2012calibrated, rana2019learning, rana-icme17, rana-icip17}, aesthetics image analysis~\cite{datta2006studying, aydin2015automated}, color-reproduction in cameras~\cite{Karaimer_CVPR},
image/video quality assessment~\cite{winkler2012analysis, pinson2013selecting,yendrikhovskij1998optimizing, panetta2013no}, virtual reality~\cite{rana-icassp} etc. For colorfulness estimation, several methods have been proposed~\cite{yendrikhovskij1998optimizing, hasler2003measuring, datta2006studying, panetta2013no, amati2014study} in the literature, in addition to the techniques used in the color appearance models~\cite{hunt1995reproduction, nayatani1995revision, fairchild2013color} (see Section~\ref{sec:relatedWork} for more discussion).  


To explore the competence of learning-based approaches and to create a stepping stone for further analysis of human perception with deep learning methods, in this paper, we propose a novel deep learning-based objective metric `ColorNet' for the estimation of colorfulness in natural images. 
Based on a convolutional neural network (CNN), our proposed ColorNet is a two-stage color rating model, where at stage I, a feature network extracts the characteristics features from the natural images and at stage II, a rating network estimates the colorfulness rating. To design our feature network, we explore the designs of the popular high-level CNN based feature models such as VGG~\cite{Simonyan15}, ResNet~\cite{He2016DeepRL}, and MobileNet~\cite{MobileNetsEC} architectures which we finally alter and tune for our colorfulness metric problem at hand. We also propose a rating network which is simultaneously learned to estimate the relationship between the characteristic features and ideal colorfulness scores.  

In this paper, we additionally overcome the challenge of the absence of a well-annotated dataset for training and validating ColorNet model in a supervised manner. To this end, we combine two publicly available colorfulness databases~\cite{hasler2003measuring,amati2014study} using the results of a new subjective test which employs a common subset of both databases. The resulting `Combined' dataset contains 180 color images and corresponding subjective colorfulness scores. Finally, we compare and showcase how our ColorNet model outperforms the state-of-the-art traditional colorfulness estimation models both quantitatively and qualitatively.


\begin{figure*}[!htb]
    \centering
    \subfigure[Subset$_{\text{EPFL}}$]{
    \includegraphics[width=0.25\columnwidth]{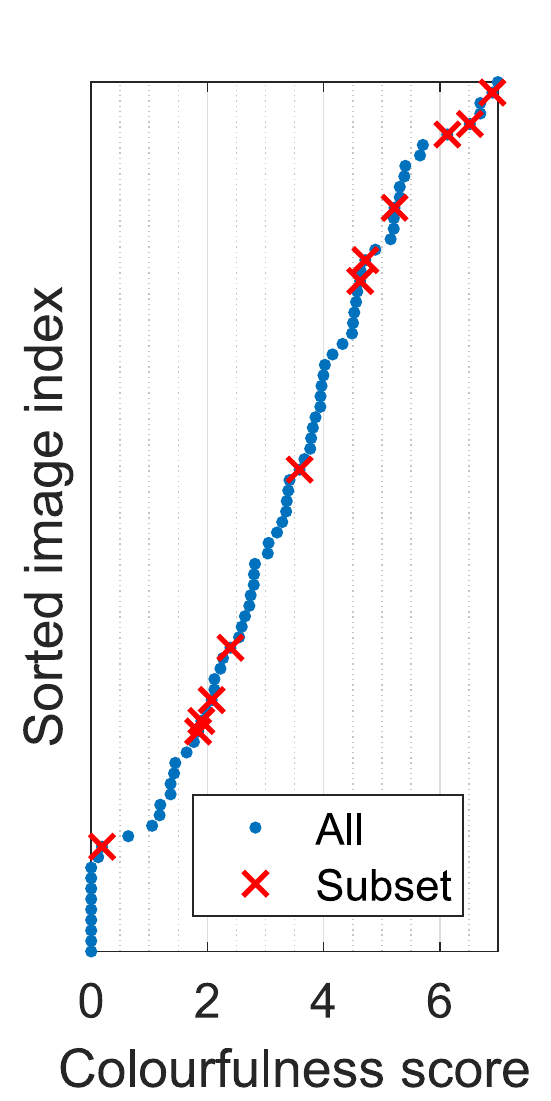}}
    \subfigure[EPFL vs. `Anchor']{
    \includegraphics[width=0.48\columnwidth]{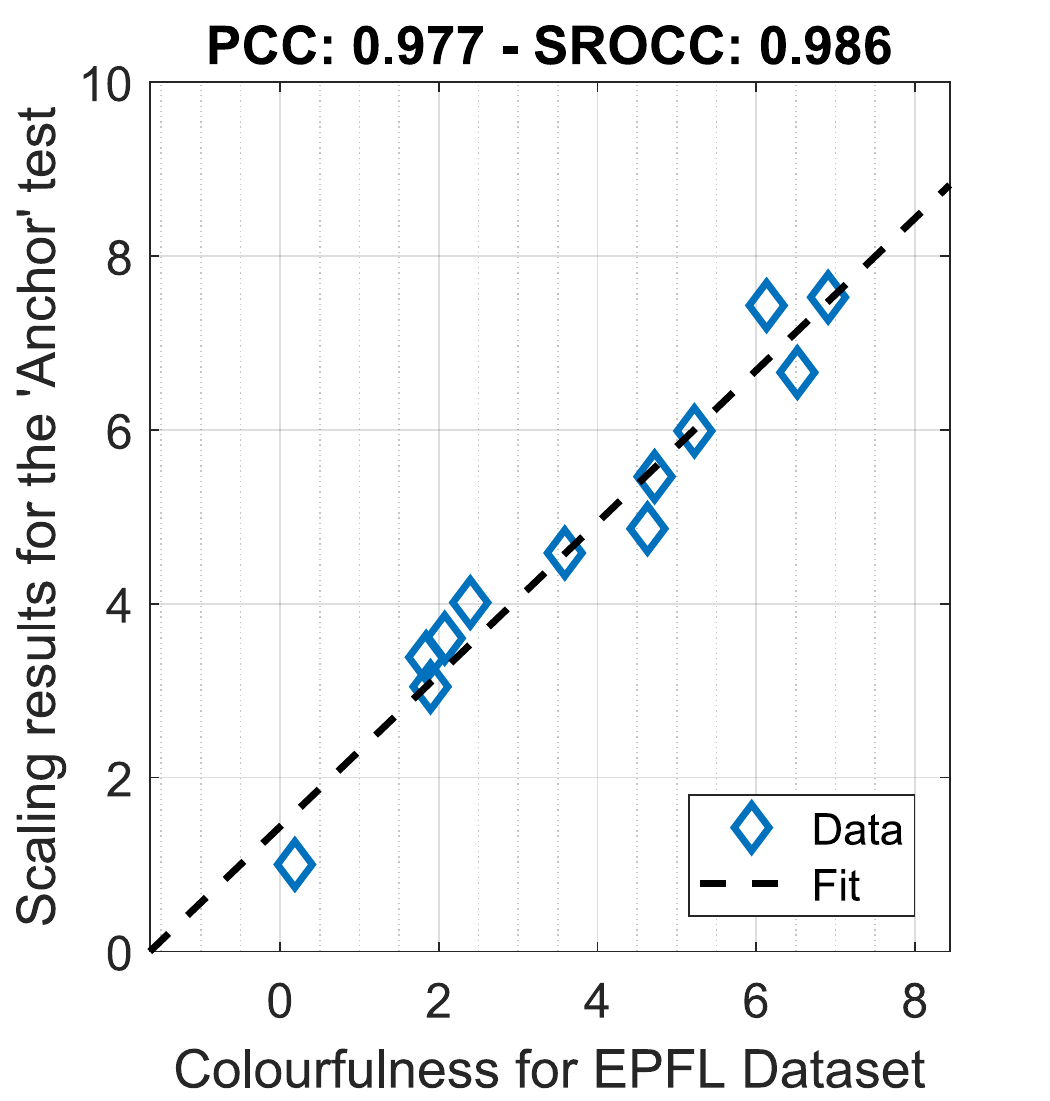}}
    \subfigure[Subset$_{\text{UCL}}$]{
    \includegraphics[width=0.25\columnwidth]{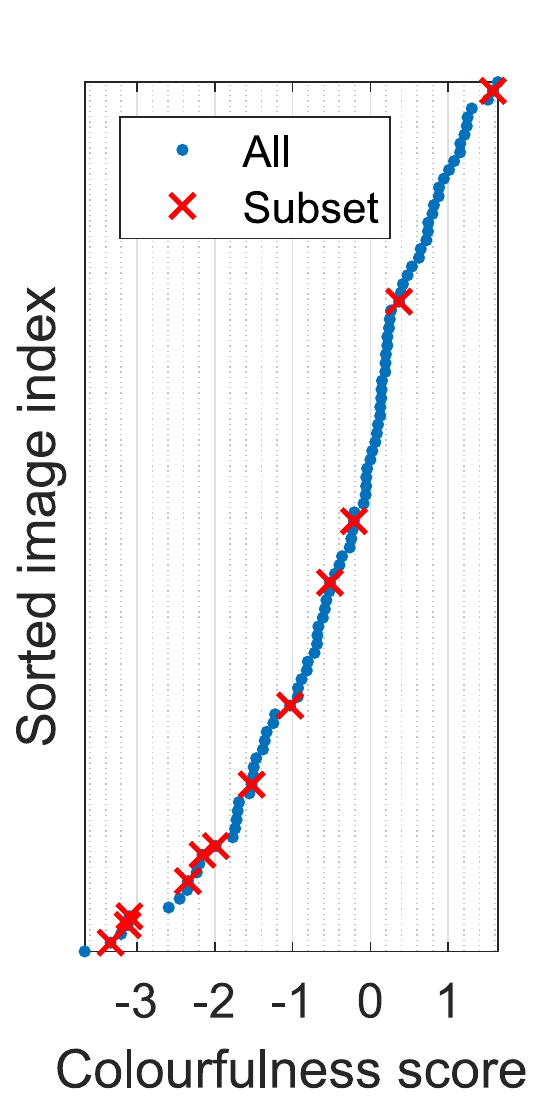}}
    \subfigure[UCL vs. `Anchor']{
    \includegraphics[width=0.48\columnwidth]{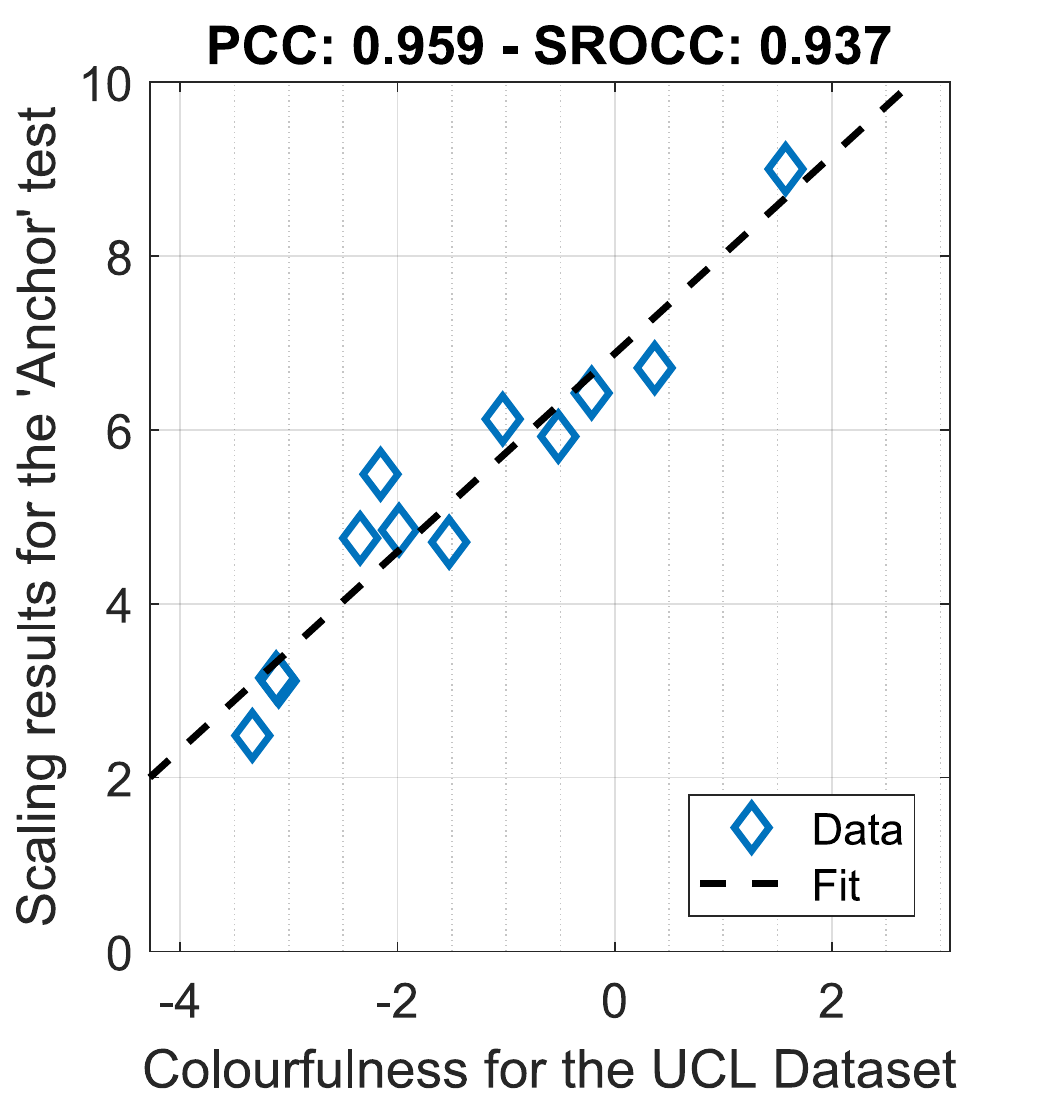}}
    \subfigure[`Combined' vs. `Anchor' ]{
    \includegraphics[width=0.48\columnwidth]{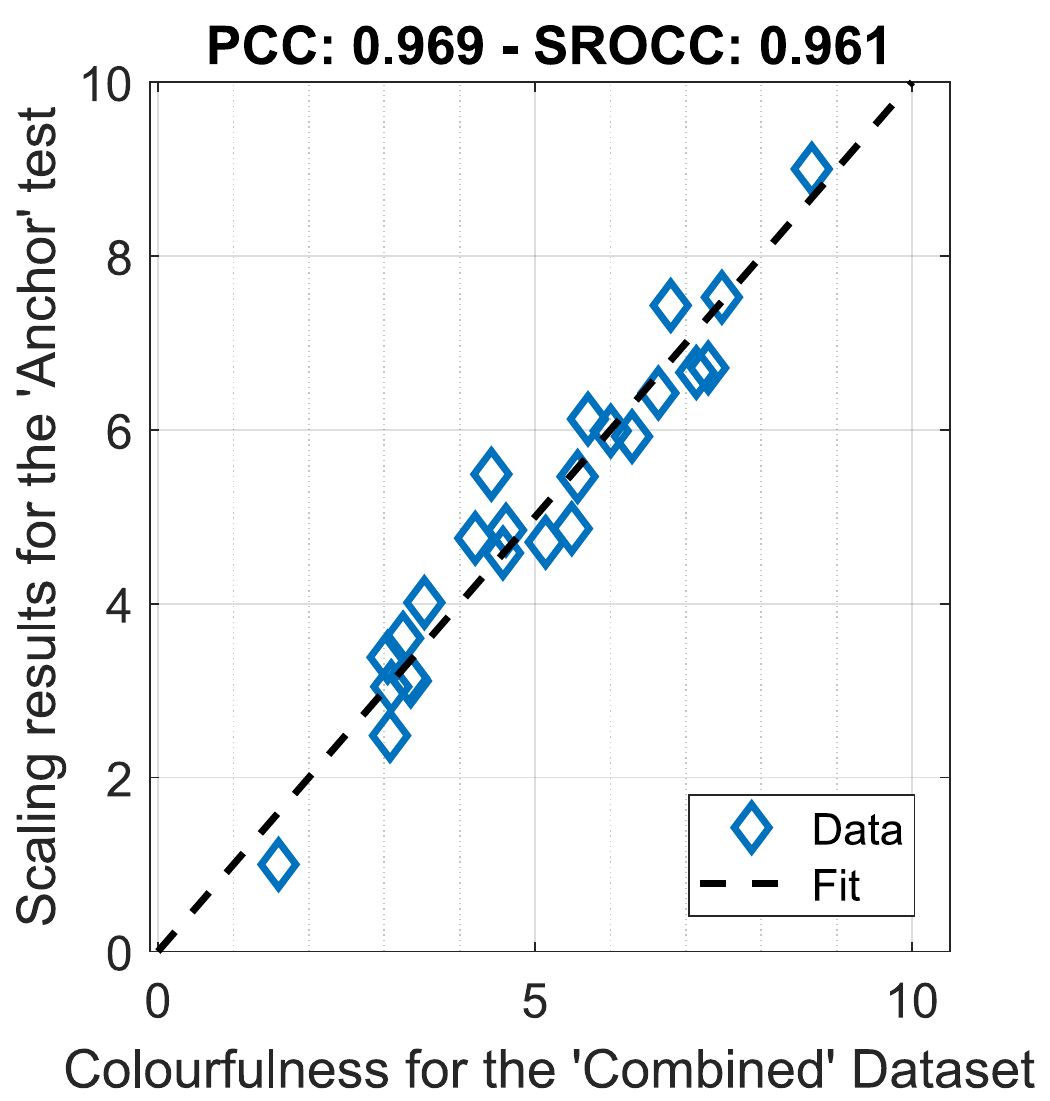}}
  \caption{Subjective data. To be representative, 12 images are selected each from EPFL~\textit{(a)} and UCL~\textit{(c)} datasets where blue circles indicate subjective colorfulness scores for all the images and red cross marks indicate those of the selected images. In (b), (d) and (e), we present the relationship between the colorfulness scores of the `Anchor' experiment conducted and the colorfulness scores of EPFL~\textit{(b)}, UCL~\textit{(d)}, and `Combined'~\textit{(e)} datasets. The diamonds indicate the scores for the selected subset images and the black dashed line indicates the best linear fit between the considered two scores.}
  \label{fig:subjScores}
  \vspace{-4pt}
\end{figure*}

\section{Related Work}
\label{sec:relatedWork}

Several studies have attempted to understand and estimate the colorfulness in visual content. Color appearance models (CAMs) utilize some form of chroma and colorfulness estimation to estimate the local visual perception and to reproduce the colors considering the lighting conditions~\cite{nayatani1995revision, hunt1995reproduction, fairchild2013color}. However, such estimations are valid mostly for simple uniform image patches. To estimate the overall colorfulness values of complex natural images, several studies have been conducted~\cite{yendrikhovskij1998optimizing, hasler2003measuring, datta2006studying, amati2014study, panetta2013no} in the literature. 

Yendrikhovskij et~al.~\cite{yendrikhovskij1998optimizing} developed a model to estimate the quality of natural color images by computing their naturalness and colorfulness indices. The proposed colorfulness index of an image is given as a summation of the average saturation (saturation as defined in CIE L*u*v* color space) and the standard deviation of the saturation. In another study, Datta et~al.~\cite{datta2006studying} proposed to first partition the RGB cube and estimate the colorfulness by calculating the Earth Mover's Distance between the pixel distribution of the tested image and that of an ``ideal'' colorful image. However, compared to others, the model failed in perceptual validation~\cite{amati2014study}.

Hasler and S\"usstrunk~\cite{hasler2003measuring}, on the other hand, studied the colorfulness in natural images by conducting a subjective user study over a relatively complex set of images. Their proposed method estimates the colorfulness by computing the first and second order statistics between the opponent color channels \ie yellow-blue and red-green color channels of the given RGB image. Using the same opponent color space strategy, Panetta et~al.~\cite{panetta2013no} proposed a colorfulness metric as a part of their no-reference image quality measure. Their proposed model involved statistical computations in logarithmic space assuming that human visual perception works logarithmically.


The performances of these colorfulness methods have been compared in a couple of studies~\cite{amati2014study, krasula2014objective}. Amati et~al.~\cite{amati2014study} analyzed the relationship between the colorfulness and the aesthetics of the images by conducting a subjective study with 100 images. They also proposed a contrast-based colorfulness metric which, however, was not better than Hasler and S\"usstrunk~\cite{hasler2003measuring}. Hence, it was not considered in this study.


Considering a tone mapping scenario, Krasula et~al.\cite{krasula2014objective} compared three different colorfulness methods (namely $CQE^{CF}_1$, $CQE^{CF}_2$~\cite{panetta2013no}, and $CIQI_c$ --a colorfulness metric inspired from Hasler and S\"usstrunk~\cite{hasler2003measuring}) along with three naturalness and six contrast metrics. To this end, they employed the high dynamic range (HDR) image tone mapping methods database of {\v{C}}ad{\'\i}k et~al.~\cite{cadik2008evaluation} with 42 images. As a result of this comparison, the $CQE^{CF}_1$ metric was found to be the most consistent and most correlated colorfulness metric. 

In addition to the traditional image processing methods, several deep learning techniques have been recently explored for designing learning-based image quality metrics~\cite{Talebi2018NIMA, Kangcvpr}. An early attempt was made by~\cite{Kangcvpr} for no-reference image quality assessment task where the efficacy of utilizing the high-level CNN features was explored. In~\cite{Talebi2018NIMA}, a neural metric has been proposed to predict the aesthetic and pixel-level quality distributions. However, no work has been done to estimate colorfulness in images.  
In this study, we, therefore, first gather a 180-image dataset with subjective colorfulness scores and then propose a color quality estimation model by exploring various state-of-the-art high-level CNN features. 


\section{Subjective Data}
\label{sec:dataset}

In this study, we use the subjective colorfulness scores collected from participants for two different colorfulness databases: \textit{EPFL Dataset}~\cite{hasler2003measuring} with 84 images and \textit{UCL Dataset}\footnote{University College London Colourfulness Dataset - \url{http://reality.cs.ucl.ac.uk/projects/image-colourfulness/image-colourfulness.html}}~\cite{amati2014study} with 96 images\footnote{Although the links are provided for 100 images the UCL Dataset, four out of these 100 images have been removed from Flickr.}. 

Both datasets provide the collected subjective scores and corresponding images. The UCL Dataset provides pairwise comparison scores (also with the user confidence). The numerical colorfulness scores, for this database, are obtained through a Thurstone Case V scaling\footnote{\texttt{pwcmp} software -- \url{https://github.com/mantiuk/pwcmp}}~\cite{perezOrtiz2017practical}. The quality scores for the EPFL Dataset are already scaled by the respective authors, using one of the methods proposed by Engeldrum~\cite{engeldrum2000psychometric}.

Even though a psychometric scaling algorithm has been employed, the EPFL Dataset scores have been collected using a rating methodology. Whereas, UCL Dataset scores have been collected using a ranking (i.e. pairwise comparison) methodology. To bring these scores to the same scale, a third subjective test is conducted as an anchor, using a common subset of these two datasets. The two databases are then combined (i.e. aligned) using the scores from this third subjective experiment~\cite{pitrey2011aligning} as explained below.

\textbf{Selection of Images from Each Dataset}: 
To create a common subset, we first selected 12 representative images from each database. These selected images cover the whole quality scale, starting from not colorful at all to very colorful. The distribution of subjective colorfulness scores for all of the images and the selected images is shown in 
Fig.~\ref{fig:subjScores}.(a) and Fig.~\ref{fig:subjScores}.(c) 
for EPFL and UCL, respectively. These selected images are used in a new subjective experiment. The new subjective test and the subjective scores collected are referred to as `\textit{Anchor}' experiment and dataset, respectively.

\textbf{Subjective Test for Anchoring the Two Datasets}: 
To merge EPFL and UCL datasets on the same scale, we have conducted a third subjective experiment with a common subset of images~\cite{pitrey2011aligning}. The pairwise comparisons (PWC) methodology is chosen in order to keep the cognitive load for the participants lighter and the experiment process easier. Although PWC is easier for subjects to decide, the full PWC design (with $n(n-1)/2$ pairs) is time-consuming. Instead, in this study, we use an adaptive pair selection process, known as adaptive square design (ASD)~\cite{li2013boosting, P915}. With this pair selection process (and also test methodology), the pair selection and scaling are done iteratively, making sure that similar stimuli are compared more than different stimuli (for which the difference will be clearer with fewer comparisons). The ASD is implemented in Matlab, and a Thurstone Case V based psychometric scaling method is used for this experiment. 
For the presentation and interactive voting, a Matlab toolbox called Psychtoolbox is used~\cite{kleiner2007whats}.


To obtain the subjective scores, the 24 selected images are used in the `Anchor' experiment, to which three expert viewers have attended. Instead of updating the rank matrix of ASD for each observer, a different approach is used. The rank matrix of ASD is randomly initialized for each expert observer, and throughout the 5 loops of the test, this rank matrix is updated, and the new pairs are selected considering this updated rank matrix. This ensured that different images have been compared for each different expert viewer. 

The obtained PWC results are then scaled to quality values which are later used to merge the databases. The scaling results are generally arbitrary (\ie relative to each other without an origin point), and hence, hard to understand. Therefore, these subjective quality scores are mapped to [1 9] scale, 1 being the least colorful and 9 being the most colorful. 


\textbf{Combining the Datasets}: 
The relationship between the new `Anchor' colorfulness scores and those of EPFL and UCL are shown in 
Fig.~\ref{fig:subjScores}.(b) and~\ref{fig:subjScores}.(d), 
respectively, for the selected images. The relationship between these scores is linear, with very high correlation scores. 

To merge the databases linearly with corresponding 'Anchor' scores, the parameters $a$ and $b$ are found by solving $y = ax + b$ relationship. Here, $y$ is the `Anchor' score and $x$ is the source (either EPFL or UCL) database score. These parameters are found as: $a_{EPFL}=0.8748$  and  $b_{EPFL}=1.4350$ for EPFL and $a_{UCL}=1.1388$  and  $b_{UCL}=6.8759$ for UCL databases.


Then, the two databases are brought to the same scale as:
\begin{equation}
    \widehat{Q_{DB}} = a_{DB} \times Q_{DB}  + b_{DB}
\end{equation}
Finally, the mapped quality scores are concatenated in a single dataset. In Fig.~\ref{fig:subjScores}.(e), the merged (denoted as `Combined') subjective quality scores are plotted again vs the `Anchor' scores to validate the merging operation, where these two databases are observed to be on the same scale.

\section{Proposed Color Rating Model}
\label{sec:model}
\begin{figure}[tb]
    \centering
    \includegraphics[width=\columnwidth]{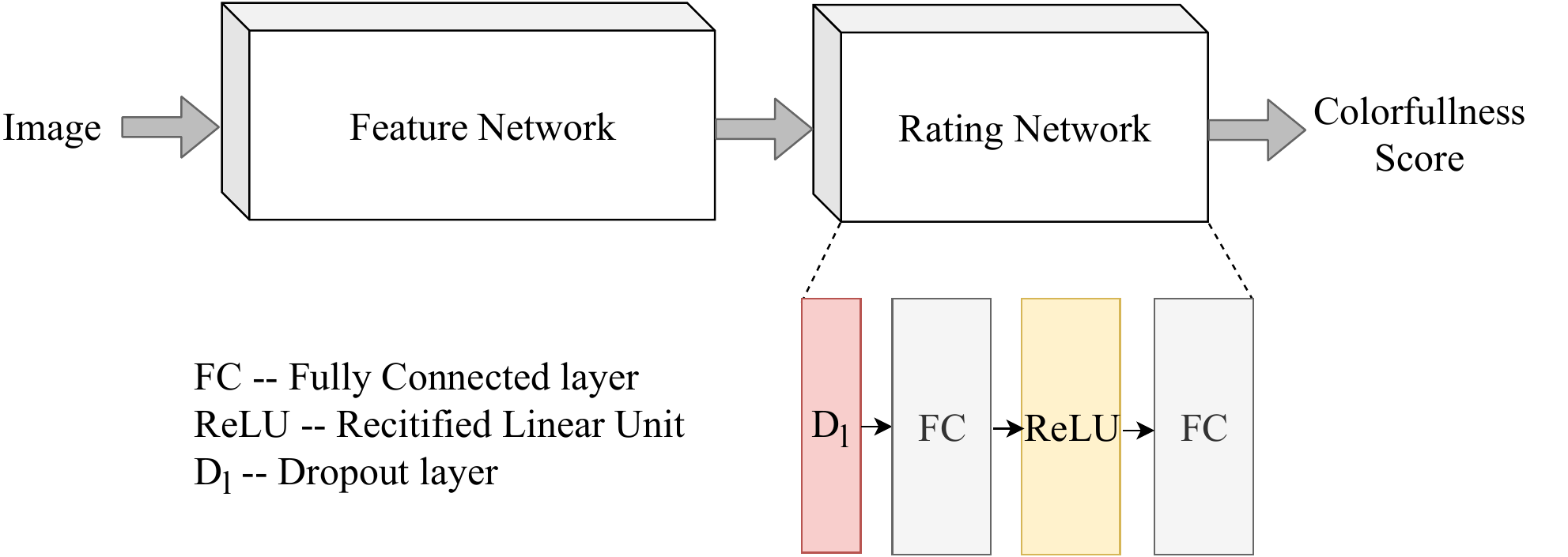}
  \caption{The ColorNet Model.}
  \label{fig:model}
  \vspace{-4pt}
\end{figure}

The ColorNet architecture is illustrated in Fig.~\ref{fig:model}. The proposed model has two major building blocks, \textit{1)} the feature network $\Phi_f$ and \textit{2)} the rating network $\Phi_r$. We base our feature network $\Phi_f$ on the state-of-the-art deep learning models, namely, VGG~\cite{Simonyan15}, ResNet~\cite{He2016DeepRL} and MobileNet~\cite{MobileNetsEC}. For all these models, we specifically removed the last layers originally meant for classification and fine-tuned the remaining feature layers in an end-to-end fashion. These features are fed to the rating network $\Phi_r$ which has an objective of mapping the characteristic features to the colorfulness rating domain. The three variants of our ColorNet model are named as \textit{ColorNet-VGG}, \textit{ColorNet-ResNet}, and \textit{ColorNet-Mobile}. 
The rating network is the same for all three variants. It consists of a dropout regularization layer, two fully connected layers $FC-10$ and $FC-1$ with $10$ and $1$ channels respectively. An added non-linearity is introduced by using the ReLU unit in between the fully connected layers to learn the desired colorfulness mapping. Further details regarding the three variants are as follows:
\begin{enumerate}
    \item \textit{ColorNet-VGG} has a 13 convolutional layered feature network $\Phi_f^{vgg}$ adopted from the VGG16~\cite{Simonyan15} architecture, with small $3\times3$ convolutions, resulting in a feature vector of dimension $512\times7\times7$. In our paper, we removed the three fully connected layers of the VGG16 architecture and simply fed the resulting feature vector into the proposed rating network.
    
    \item \textit{ColorNet-ResNet} consists of an 18 layer deep residual feature network $\Phi_f^{resnet}$ adopted from the ResNet architecture~\cite{He2016DeepRL}. Similar to VGG~\cite{Simonyan15}, the ResNet architecture has small $3\times3$ convolutions, however with an additional concept of residual learning applied to every few stacked layers. In our paper, we removed the last fully connected layer of 1000 channels and fed the resulting feature of size $512\times7\times7$ into the rating work. 
    
    \item \textit{ColorNet-Mobile} consists of a 28 convolution layers feature network $\Phi_f^{mobile}$ adopted from the MobileNet architecture~\cite{MobileNetsEC}, which comprises of both depth-wise and point-wise convolutions. MobileNet has been a widely adopted network for many mobile and embedded applications. In this paper, we removed its last fully connected and soft-max layer to obtain a feature of size $1024\times7\times7$ which is finally fed into the rating network.
\end{enumerate}

Note that we choose these feature models mainly due to ease in training and faster convergence over small dataset size. For further details regarding the feature network architectures, we refer the reader to~\cite{Simonyan15,He2016DeepRL,MobileNetsEC}. For each ColorNet variant, the receptive input for the first fully-connected is altered as per the resulting feature size of the feature network. 

\textbf{Loss Function}: The ultimate goal of ColorNet models is to rate the colorfulness in images. To this end, we train these models in an end-to-end fashion using a fully supervised setting. For training, we define a set of an image and its corresponding rating as a pair, given as $(X_i,y_i)$, where $X_i$ is an image and $y_i$ is its corresponding colorfulness rating, $i = [1,n]$, $n$ = number of training image pairs. To train the ColorNet models, we used the following objective function:
\begin{equation}
    \mathcal{L}^{j}(y,\hat{y}) = \frac{1}{n}\sum_{k=1}^{n} |y_k - \phi_r^{j}.\phi_f^{j}(X_i)|
\end{equation}
where $\hat{y}$ is the predicted colorfulness rating and $j = \{$vgg, resnet, mobile$\}$. We additionally experimented with the L2 norm as a loss function, however, we observe that practically the L1 loss term effectively penalizes the network to learn and converge at a faster rate using the ADAM~\cite{KingmaB14adam2} optimizer technique. 


\section{Results}
\label{sec:results}
\textbf{Training and Implementation Details}: 
We split the dataset of $180$ images into training, validation and test sets in an $80\%$, $10\%$ and $10\%$ setting. For training, the input image size is fixed at $600\times600$, and random crops of size $512\times512$ are applied to the image. Additional data augmentation techniques such as rotation and flipping are applied to scale up the training dataset. 

The ColorNet model is implemented using the Pytorch~\cite{paszke2017automatic} deep learning library. During training, the batch size is set to 4 and the baseline weights of the feature networks are initialized by training on the ImageNet~\cite{imagenet_cvpr09} dataset. The weights of the layers in the rating networks are initialized randomly. In other terms, we fine-tune the feature network and train the rating network layers for our task at hand. A dropout rate of 0.75 is set in the rating network for all the three models. We utilize an ADAM solver~\cite{KingmaB14adam2} with an initial learning rate of $1 \times 10^{−4}$ and $1 \times 10^{−3}$ for training the feature and rating networks respectively. The learning rates are allowed to decay exponentially with a decay rate of $0.95$, after every 10 epochs. The momentum rate is fixed at $0.9$ for all epochs. All our models are trained in an end-to-end fashion for $200$ epochs. 
Training is done using a $12$ GB NVIDIA Titan-X GPU on an Intel Xeon E7 core i7 machine for $200$ epochs which take approximately $2$ hours. Inference time is $25$ secs for each image. 

\textbf{Traditional Colorfulness Methods}: 
We implemented four different colorfulness metrics: Hasler and S\"usstrunk ($CF_\text{Hasler}$)~\cite{hasler2003measuring} , two versions ($CQE^{CF}_1$ and $CQE^{CF}_2$) of Panetta et~al.~\cite{panetta2013no}, and Yendrikhovskij et~al.  ($CF_\text{Yendrikhovskij}$)~\cite{yendrikhovskij1998optimizing} to compare with the proposed ColorNet model. 

For the Hasler and S\"usstrunk~\cite{hasler2003measuring}, with a given RGB image, the metric uses the mean $\mu$, and standard deviation $\sigma$ of the opponent color space vectors $v_{rg}$ and $v_{yb}$ where $v_{rg} = I_R - I_G$ and $v_{yb} = (I_R + I_G)/2 - I_B$. Then, $CF_\text{Hasler}$ is computed as:
\begin{equation}
\label{eqn:haslerCF}
    CF_\text{Hasler} = \sqrt{\sigma_{rg}^2 + \sigma_{yb}^2} + 0.3 \times \sqrt{\mu_{rg}^2 + \mu_{yb}^2}
\end{equation}

Panetta et~al.~\cite{panetta2013no} use the similar color opponent space, but in logarithmic space. The metrics $CQE_{1}^{CF}$ and $CQE_{2}^{CF}$ are calculated as:
\begin{align}
\label{eqn:cqeCF}
\begin{split}
    CQE_{1}^{CF} &= 0.02 \times \log\left(\frac{\sigma_{rg}^2}{|\mu_{rg}|^{0.2}}\right) \times \log\left(\frac{\sigma_{yb}^2}{|\mu_{yb}|^{0.2}}\right)
\\
    CQE_{2}^{CF} &= 0.02 \times \frac{\log(\sigma_{rg}^2) \times \log(\sigma_{yb}^2)}{\log(\sigma_c^2)} \times \frac{\log(\mu_{rg}^2) \times \log(\mu_{yb}^2)}{log(\mu_c^2)}
\end{split}
\end{align}
where $v_{rg}$ and $v_{yb}$ are concatenated to form $v_{c}$, \ie $v_{c} = [v_{rg}, v_{yb}]$. 

To compute $CF_\text{Yendrikhovskij}$, we use the following formula after the RGB to CIE L*u*v* color space transform, as specified in the paper~\cite{yendrikhovskij1998optimizing}: 
\begin{align}
\label{eqn:yendrikCF}
\begin{split}
    S_{u^{*}v^{*}} &= \frac{ \sqrt{ ({{u^{*}}^2 + {v^{*}}^2}) } }{L^{*}+\epsilon}, \epsilon \neq 0
\\
    CF_\text{Yendrikhovskij} &= \mu_{S_{u^{*}v^{*}}} + \sigma_{S_{u^{*}v^{*}}}
\end{split}
\end{align}


\begin{table}[bp]
    \vspace{-10pt}
    \centering
    \caption{Correlation Coefficient Results.} 
    \vspace{5pt}
    \begin{tabular}{l|cc} \hline \hline \noalign{\smalllskip}
        \textbf{Colorfulness Metric} 
           & \textbf{PCC} & \textbf{SROCC} \\ \noalign{\smalllskip} \hline \noalign{\smalllskip}
        $CF_\text{Hasler}$~\cite{hasler2003measuring} 
           & 0.841 & 0.884 \\
        $CQE_{1}^{CF}$~\cite{panetta2013no} 
           & 0.895 & 0.896 \\
        $CQE_{2}^{CF}$~\cite{panetta2013no} 
           & 0.312 & 0.415 \\
        $CF_\text{Yendrikhovskij}$~\cite{yendrikhovskij1998optimizing}
           & 0.843 & 0.834 \\ \noalign{\smalllskip} \hline \noalign{\smalllskip}
        ColorNet-Mobile
           & 0.841 & 0.774 \\
        ColorNet-ResNet
           & 0.916 & 0.889 \\
        ColorNet-VGG
           & \textbf{0.937} & \textbf{0.921} \\\noalign{\smalllskip} \hline \hline
    \end{tabular}
    \label{tab:corr}
\end{table}

\textbf{Quantitative Evaluation}: 
Colorfulness metrics are evaluated by computing the Pearson correlation coefficient (PCC) and the Spearman rank-ordered correlation coefficient (SROCC). For testing, we used the 10-fold cross-validation strategy where the whole dataset was divided into 10 non-overlapping pieces $\mathcal{P}$, and for each iteration, one piece is used for the test, one piece used for validation, and the remaining pieces are used for training (as also described above). For each iteration $itr$, the unique piece $\mathcal{P}_{itr}$ is used for the test, where all the pieces ($\mathcal{P}_{itr}$, $itr \in [1,10]$) cover the whole dataset. Finally, the PCC and SROCC results for 10 different test cases are averaged. The performance results are presented in Table~\ref{tab:corr} where PCC and SROCC for the 4 state-of-the-art colorfulness metrics are computed using the aligned subjective scores from `Combined' dataset. Our proposed ColorNet model with VGG and ResNet based feature network outperforms over the other classical models. This is partly due to the model's ability to learn rich high-level color feature representation. 

\begin{figure}[tb]
    \centering
    \includegraphics[width=\columnwidth]{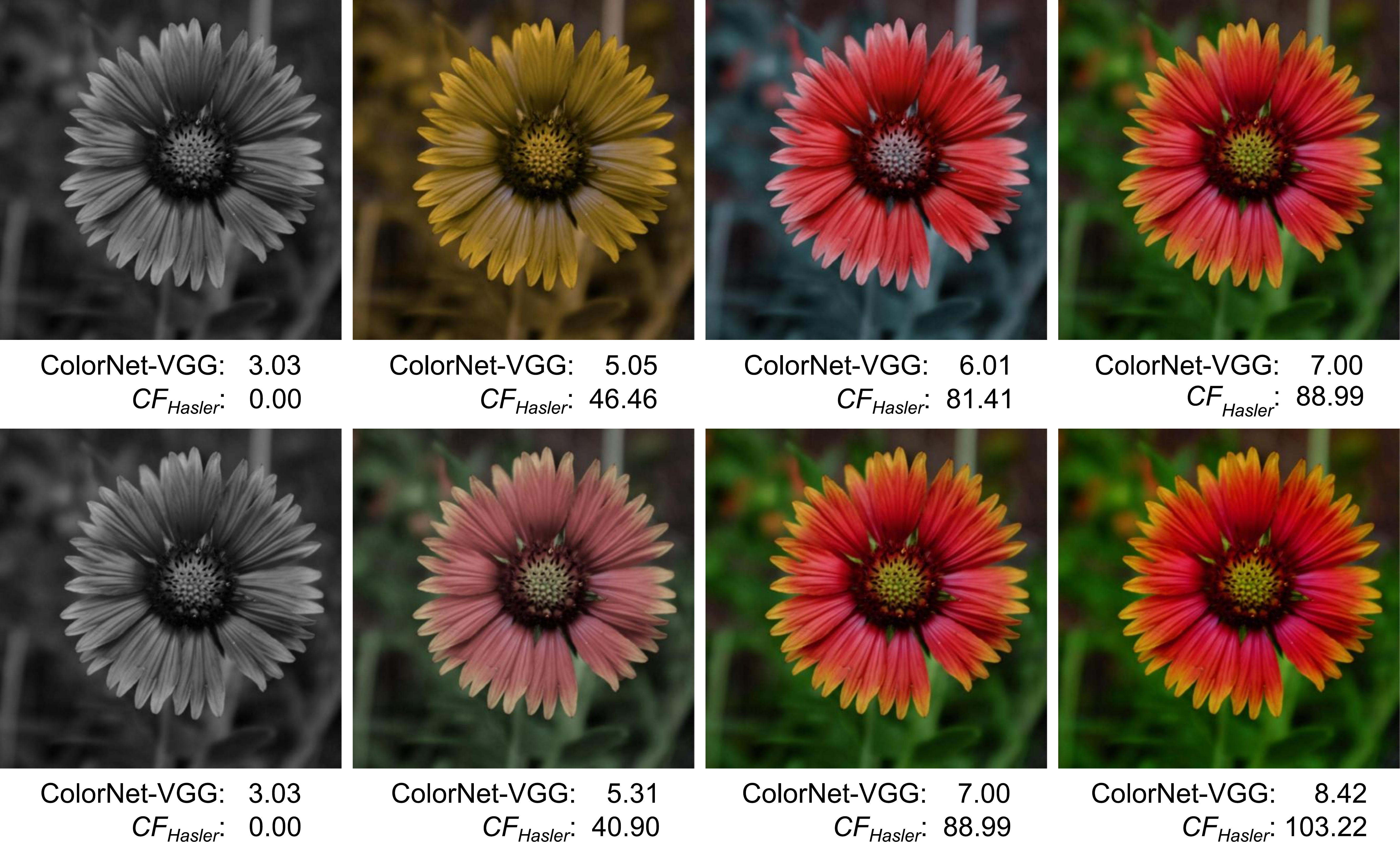}
  \caption{Qualitative Evaluation. Row I depicts the change in dominant colors. Row II depicts the change in the saturation of the colors.}
  \label{fig:flowers}
  \vspace{-4pt}
\end{figure}

\textbf{Qualitative Evaluation}:
For the qualitative evaluation of ColorNet, we crafted a set of images that are not used during the training of the model, by considering two different scenarios: \textit{i)} change in the dominant colours and \textit{ii)} change in the color contrast. In Fig.~\ref{fig:flowers}, row I shows 4 different cases of the change in the dominant colors, and row II shows the change in the color contrast. We report the objective results for the proposed deep-learning-based colorfulness estimation method, in Fig.~\ref{fig:flowers}, and show also $CF_{Hasler}$ for completeness. The results showcase that in both cases \ie by increasing color contrast and increasing the number of dominant hues, the proposed metric scores increase, thus, validating the understanding of colorfulness of our model. Overall, our results confirm that learning-based models bring huge potential to cater for various implicit aspects of color perception over a wide variety of natural images.  

\section{Conclusion}
\label{sec:conclusion}

In this study, we propose a CNN based model for the estimation of colorfulness ratings. To prepare a well-annotated colored image dataset, we combine two colorfulness databases with subjective user scores, using the results of an anchor subjective experiment with a common subset of images. We compare the results of the proposed model to those of four other traditional colorfulness metrics quantitatively and qualitatively where we observe that our learning-based model effectively rates the colorfulness by catering for the wide variety of natural images. This study constitutes an initial step towards the exploration of color perception in natural images using the deep learning approach. In future work, we aim to delve deeper in learning-based color perception models and analyze the impacts of various associated factors. 

\footnotesize
\bibliographystyle{IEEEtran}
\bibliography{refs}

\end{document}